\documentclass[12pt,prd,aps,epsfig,floats,preprint,eqsecnum]{revtex4}
\usepackage{graphicx}

\newcommand{\be}{\begin{equation}}
\newcommand{\ee}{\end{equation}}
\newcommand{\bea}{\begin{eqnarray}}
\newcommand{\eea}{\end{eqnarray}}
\newcommand{\beq}{\begin{equation}}
\newcommand{\eeq}{\end{equation}}
\newcommand{\beqa}{\begin{eqnarray}}
\newcommand{\eeqa}{\end{eqnarray}}
\newcommand{\no}{\nonumber}
\def\lsim{\mathrel{\rlap{\lower4pt\hbox{\hskip1pt$\sim$}}
     \raise1pt\hbox{$<$}}}         %less than or approx. symbol
\def\gsim{\mathrel{\rlap{\lower4pt\hbox{\hskip1pt$\sim$}}
     \raise1pt\hbox{$>$}}}         %greater than or approx. symbol
\begin{document}
%\draft

%\preprint{{\vbox{\hbox{}\hbox{}\hbox{}
\hbox{DO-TH-08/07}
\hbox{WIS/21/08-Nov-DPP}
%\hbox{hep-ph/yymmnnn}}}}

\vspace*{.0cm}

\title{Flavor Changing Processes in Supersymmetric Models\\
with Hybrid Gauge- and Gravity-Mediation}

\author{Gudrun Hiller}\email{ghiller@physik.uni-dortmund.de}
\affiliation{Institut f\"ur Physik, Technische Universit\"at Dortmund,
   D-44221 Dortmund, Germany}
\author{Yonit Hochberg and Yosef Nir\footnote{The Amos de-Shalit chair of theoretical
     physics}}\email{yonit.hochberg,yosef.nir@weizmann.ac.il}
\affiliation{Department of Particle Physics,
   Weizmann Institute of Science, Rehovot 76100, Israel\vspace*{1cm}}

\vspace*{1cm}
\begin{abstract}
We consider supersymmetric models where gauge mediation provides the
dominant contributions to the soft supersymmetry breaking terms
while gravity mediation provides sub-dominant yet non-negligible
contributions. We further assume that the gravity-mediated
contributions are subject to selection rules that follow from a
Froggatt-Nielsen symmetry. This class of models constitutes an
example of viable and natural non-minimally flavor violating models.
The constraints from $K^0-\overline{K}^0$ mixing imply that the
modifications to the Standard Model predictions for
$B_d-\overline{B}_d$ and $B_s-\overline{B}_s$ mixing are generically at most at
the percent level, but can be of order ten percent for large $\tan \beta$.
The modifications for $D^0-\overline{D}^0$
mixing are generically at most of order a few percent, but in a
special subclass of models they can be of order one. We point
out $\Delta B=1$ processes relevant for flavor violation
in hybrid mediation.
\end{abstract}
\maketitle

%%%%%%%%%%%%%%%%%%%%%%%%%
\section{Introduction}

The physics of flavor and CP violation could be rich with deviations
from the Standard Model  predictions if supersymmetry is realized
at the TeV scale, and if the mechanism that mediates its breaking to
the minimal supersymmetric standard model (MSSM) is not minimally
flavor violating (MFV). Indeed, hybrid models of gauge- and
gravity-mediation can lead to flavor violating effects large enough to be
explored by the LHC experiment \cite{Feng:2007ke,Nomura:2007ap}. It is
the purpose of this work to study whether the flavor- and CP-violating
effects expected in this framework can be discovered in decays of $D$,
$B_d$ and $B_s$ mesons.

The basic idea of the hybrid gauge-gravity models is the following.
There are gauge-mediated contributions to the soft supersymmetry
breaking terms. If the scale of gauge mediation is low, then the gravity-mediated
contributions are negligible, and the model is MFV. If the scale is
very high, then the gravity-mediated contributions dominate, and it
requires a very careful model building to suppress the supersymmetric
contributions to flavor changing neutral current (FCNC) processes
\cite{Nir:2002ah}. There is, however, an intermediate range for the
scale of gauge mediation where the gravity-mediated contributions are
neither negligible nor dominant \cite{ArkaniHamed:1997km}. For this
range of scales, the Froggatt-Nielsen (FN) mechanism
\cite{Froggatt:1978nt} -- an approximate horizontal Abelian symmetry
-- can play a role in suppressing the squark and slepton mixing in a
simple and natural way \cite{Nir:1993mx,Leurer:1993gy}.

If the ATLAS/CMS experiments can measure the mass splitting between
squarks or sleptons, we will learn about the relative importance of
the gauge- and gravity-mediated contributions and thereby on the
gauge mediation scale. If these experiments can measure the flavor
decomposition of squarks and sleptons, the FN framework can be tested
and we may further learn about the way that the FN symmetry is
implemented. Here we would like to ask whether the (present and
future) $B$-factories, the TeVatron experiments and the LHCb
experiment can give early hints to this framework, before direct
squark and slepton measurements are achieved.

The paper is organized as follows.  We review FCNC constraints in SUSY
models in Section \ref{sec:fcnc}. In Section \ref{sec:hybrid} we work
out the flavor-violating low energy couplings in the hybrid
gauge-gravity models, and discuss their phenomenology in view of FCNC
data in Section \ref{sec:phe}.  Section \ref{sec:hol} contains the
phenomenological consequences of a variant of FN models with
holomorphic zeros. In Section \ref{sec:messenger} we discuss general
properties of gauge mediation in the context of flavor constraints and
comment on hidden sector effects. We conclude in Section
\ref{sec:con}.  The Appendix contains details on the effects of MSSM
renormalization group running.

%%%%%%%%%%%%%%%%%%%%
\section{FCNC constraints on SUSY parameters}
\label{sec:fcnc}
%%%%%%%%%%%
New physics at the TeV scale could lead to enhancement of FCNC
processes by orders of magnitude. The fact that such an enhancement
has not been observed in any of the $s\to d$, $c\to u$, $b\to d$ and
$b\to s$ transitions gives strong constraints on the flavor structure
of the new physics. We discuss constraints on SUSY parameters
from gluino loops in Section \ref{sec:gluino} and from
chargino contributions in Section \ref{sec:chargino}.
The impact of rare decays and the constraints that arise at
large $\tan \beta$ are covered in Section \ref{sec:delF1}.

\subsection{Gluino contributions \label{sec:gluino}}

In the supersymmetric framework, the following combinations of parameters
are strongly constrained by processes involving $q_i\to q_j$ transitions:
\beq
\delta_{ij}^q=\frac{1}{\tilde m^2_q}\sum_\alpha
K_{i\alpha}^{q} K_{j\alpha}^{q*}\Delta\tilde m^2_{q_\alpha}.
\eeq
Here $K_{i\alpha}^q$ is the mixing angle in the coupling of the gluino
(and similarly the bino and neutral wino) to $q_i-\tilde q_\alpha$,
$\tilde m^2_q=\frac13\sum_{\alpha=1}^3\tilde m^2_{q_\alpha}$ is the
average squark mass-squared, and $\Delta\tilde m^2_{q_\alpha}=\tilde
m^2_{q_\alpha}-\tilde m^2_q$. Using the unitarity of the mixing matrix
$K$, we can write
\beq
 \label{eq:deltaq-def}
\tilde m^2_q \delta_{ij}^q=\sum_\alpha
K_{i\alpha}^{q} K_{j\alpha}^{q*}(\Delta\tilde m^2_{q_\alpha}+\tilde
m^2_q)=(\widetilde M^2_q)_{ij},
\eeq
where $\widetilde M^2_q$ is the mass-squared matrix for the squarks
$\tilde q$ in the basis where the quark $q$ masses and the gluino
couplings are diagonal.

The mass-squared matrices carry also chirality indices, $M,N=L,R$,
{\it i.e.} $(\widetilde M^2_q)^{MN}_{ij}$ is the $\tilde
q_{Mi}^{\dagger}\tilde q_{Nj}$ mass-squared term. Correspondingly, the
$\delta_{ij}^q$ are assigned chirality indices, namely the FCNC
constrain $(\delta_{ij}^q)_{MN}$. In the case that the $\tilde
q_L-\tilde q_R$ mixing can be neglected, there are four classes of
$(\delta_{ij}^{q})_M\equiv(\delta_{ij}^q)_{MM}$: $(\delta_{ij}^{d})_L$ for
the left-handed down squarks $\tilde D_L$, $(\delta_{ij}^{u})_L$ for the
left-handed up squarks $\tilde U_L$, $(\delta_{ij}^{d})_R$ for the
right-handed down squarks $\tilde D_R$, and $(\delta_{ij}^{u})_R$ for
the right-handed up squarks $\tilde U_R$. We also define
\beq
\langle\delta^q_{ij}\rangle=\sqrt{(\delta_{ij}^{q})_L (\delta^{q}_{ij})_R}.
\eeq

In some cases, a two generation effective framework is useful. To
understand that, consider a case where (no summations over $i,j,k$):
\beqa\label{condtwogen}
|K_{ik} K_{jk}^*|&\ll&|K_{ij} K_{jj}^*|,\no\\
|K_{ik} K_{jk}^*\Delta\tilde
m^2_{q_kq_i}|&\ll&|K_{ij} K_{jj}^*\Delta\tilde m^2_{q_j q_i}|,
\eeqa
where $\Delta\tilde m^2_{q_jq_i}=\tilde m^2_{q_j}-\tilde
m^2_{q_i}$. Then the contribution of the intermediate $\tilde q_k$ can
be neglected and, furthermore, to a good approximation,
$K_{ii} K_{ji}^*+K_{ij} K_{jj}^*=0$. For these cases, we obtain
\beq\label{twogendel}
\delta_{ij}^q=\frac{\Delta\tilde m^2_{q_jq_i}}{\tilde m^2_q}
K_{ij}^{q} K_{jj}^{q*}.
\eeq
It is further useful to use instead of $\tilde m_q$ the following
average mass scale \cite{Raz:2002zx}:
\beq
\tilde m_{ij}^q=\frac12(\tilde m_{q_i}+\tilde m_{q_j}).
\eeq

\begin{table}[t]
\caption{The phenomenological upper bounds on $(\delta_{ij}^{q})_A$ and
   on $\langle\delta^q_{ij}\rangle$, where $q=u,d$ and $A=L,R$.
   The constraints are given for $m_{\tilde q}=1$ TeV and $x\equiv m_{\tilde
   g}^2/m_{\tilde q}^2=1$. We assume that the phases could suppress the
   imaginary parts by a factor $\sim0.3$. The bound on
   $(\delta^{d}_{23})_R$ is about 3 times weaker than that on
   $(\delta^{d}_{23})_L$ (given in table). The constraints on
   $(\delta^{d}_{12,13})_A$, $(\delta^{u}_{12})_A$ and $(\delta^{d}_{23})_A$
   are based on, respectively, Refs. \cite{Masiero:2005ua},
   \cite{Ciuchini:2007cw} and \cite{Buchalla:2008jp}.}
\label{tab:exp}
\begin{center}
\begin{tabular}{cc|cc} \hline\hline
\rule{0pt}{1.2em}%
$q$\ & $ij\ $\ &  $(\delta^{q}_{ij})_A$ & $\langle\delta^q_{ij}\rangle$ \cr \hline
$d$ & $12$\ & $\ 0.03\ $ & $\ 0.002\ $ \cr
$d$ & $13$\ & $\ 0.2\ $ & $\ 0.07\ $ \cr
$d$ & $23$\ & $\ 0.6\ $ & $\ 0.2\ $ \cr
$u$ & $12$\ & $\ 0.1\ $ & $\ 0.006\ $ \cr
\hline\hline
\end{tabular}
\end{center}
\end{table}

Constraints of the form $\delta_{ij}^q\ll1$ imply that either
quasi-degeneracy ($\Delta\tilde m^2_{q_iq_j}\ll \tilde m^{q  2}_{ij}$) or
alignment ($|K^q_{ij}|\ll1$) or a combination of the two mechanisms
is at work. We use the constraints obtained in
Refs. \cite{Masiero:2005ua,Ciuchini:2007cw,Buchalla:2008jp}. They are
presented in Table \ref{tab:exp}. Wherever relevant, we allow a mild
phase suppression in the mixing amplitude, namely we
quote the stronger between the bounds on ${\cal R}e(\delta^q_{ij})$
and $3{\cal I}m(\delta^q_{ij})$.
We would like to emphasize the following points:
\begin{enumerate}
\item The bounds have a strong dependence on the average squark mass,
   scaling roughly as $m_{\tilde q}/(1$ TeV).
\item The bounds have a milder dependence on the ratio $x\equiv
   m_{\tilde g}^2/m_{\tilde q}^2$. In particular, for $x=4$, the
   bound on $(\delta^{d}_{12})_A$ ($\langle\delta^d_{12}\rangle$) is
   weakened to $0.06$ ($0.003$).
\item If we allow an arbitrarily strong suppression of
the CP violating phases, some bounds are further relaxed. For example,
with zero phase, $m_{\tilde q}=1$ TeV and $x=1$, we have
$\langle\delta^d_{12}\rangle\leq0.004$.
\item The bounds compiled in Table \ref{tab:exp} are based on
conservative estimates. At large $\tan\beta$ the bounds can be significantly stronger and are more model-dependent, see Section \ref{sec:delF1}.
\end{enumerate}

%%%%%%%%%%%
\subsection{Chargino contributions \label{sec:chargino}}
Chargino contributions could also be of interest. If $\tan\beta$ is
not very large, then for the various processes of interest the
charged higgsino contributions are suppressed by small Yukawa
couplings.  We focus then on the charged wino contributions to $d_i\to
d_j$ transitions, which involve intermediate $\tilde u_{L\alpha}$
squarks.  Now the following combination is constrained (we omit here
the chirality index $L$):
\beq
  \label{eq:deltacu}
\delta_{ij}^{cu}=\frac{1}{\tilde m^2_u}\sum_\alpha
Z_{i\alpha}^{u} Z_{j\alpha}^{u*}\Delta\tilde m^2_{u_\alpha}.
\eeq
Here $Z_{i\alpha}^u$ is the mixing angle in the coupling of the wino
to $d_i-\tilde u_\alpha$ (both `left-handed'). Note that
\beq
\tilde m^2_u \delta_{ij}^{cu}=\sum_\alpha
Z_{i\alpha}^{u} Z_{j\alpha}^{u*}(\Delta\tilde m^2_{u_\alpha}+\tilde
m^2_u)=(\widetilde M^{c2}_u)_{ij},
\eeq
where $\widetilde M^{c2}_u$ is the mass-squared matrix for the
left-handed up squarks $\tilde u_L$ in the basis where the {\it down}
quark masses and the gluino couplings are diagonal. Note that
$\delta^{cu}_{ij}\neq \delta^u_{ij}$. In particular,
\beq
Z^u=V^\dagger K^u,
\eeq
where $V$ denotes the CKM quark mixing matrix.

Consider, for example, $\delta^{cu}_{12}$ and assume that the
conditions for an effective 2-flavor framework, Eq.~(\ref{condtwogen}),
hold. Then, defining
$\sin\tilde\theta_u\equiv K_{12}^u$, we obtain
\beq
 \label{eq:deltacu-2gen}
\tilde m_u^2\delta^{cu}_{12}=\frac12\sin(2\tilde\theta_u-2\theta_c)
\Delta\tilde m^2_{u_2 u_1},
\eeq
where $\theta_c$ denotes the Cabibbo angle. On the other hand,
\beq
\tilde m_u^2\delta^{u}_{12}=\frac12\sin(2\tilde\theta_u)
\Delta\tilde m^2_{u_2 u_1}.
\eeq

Given a bound on $(\delta_{ij}^d)_L$ from gluino loops,
by $SU(2)$ symmetry there is a
corresponding bound on $\delta_{ij}^{cu}$
(see Section \ref{sec:flavormz} for details).
The latter is often stronger than the bound from chargino contributions
by approximately a factor of $(\alpha_3/\alpha_2)$,
though there is further dependence on the
gaugino masses via known loop functions.

If the mixing angles are small, Eq.~(\ref{eq:deltacu-2gen}), in general
involving arbitrary two generations, can be linearized and yields
\beq
\delta^{cu}_{ij} =\delta^u_{ij} + V_{ji}^*
\frac{\Delta \tilde m^2_{u_j u_i}}{\tilde m_u^2} .
\eeq
This decomposition is commonly used to constrain $\delta^u$ through chargino
interactions in rare processes, {\it e.g.}, \cite{Lunghi:1999uk}.
The separation into 'flavor diagonal' and $\delta^u$-induced terms,
however, is not useful in models where there are cancelations between
$K^u$ and the CKM matrix elements.

%%%%%%%%%%%
\subsection{$\Delta B=1$ processes and large $\tan \beta$ \label{sec:delF1}}
A multitude of $\Delta B=1$ decay observables has been measured so
far \cite{Barberio:2008fa}. The most interesting ones for the
purpose of constraining new physics parameters are those which have
reasonable theoretical and experimental uncertainties, and depend only
on a small set of model parameters. Given these requirements, very
useful modes are radiative and (semi)-leptonic decays mediated by $b
\to q \gamma$ and $b \to q \ell^+ \ell^-$ for $\ell=e,\mu$ and
$q=d,s$.  Currently, theory gives preference to inclusive versus
exclusive decays, although the purely leptonic and very rare $B \to
\ell^+ \ell^-$ decays are also important. Future data on dedicated
distributions and asymmetries in FCNC exclusive decays, which will
become available in the LHC era \cite{Buchalla:2008jp}, will also be
of relevance.

For the constraints on $\delta_{23}^d$ in Table \ref{tab:exp}, data on
$B \to X_s \ell^+ \ell^-$, $B \to X_s \gamma$ decays and $B_s$ mixing
has been employed.  For the radiative and semileptonic $b \to d$
decays, the experimental situation is currently not as good as
for $b \to s$ decays, and only $B_d$ mixing has been used to limit
$\delta^d_{13}$.

The impact of $\Delta B =1$ versus $\Delta B=2$ processes for the
bounds on the $\delta^q$ parameters has a complex dependence on the
model parameters.  For example, for $(\delta^d_{23})_R$, the strongest
constraint comes from $\Delta B=2$, whereas for $(\delta^d_{23})_L$
and $\langle \delta^d_{23} \rangle$ the rare $B \to X_s \gamma$
and $B \to X_s\ell^+\ell^-$ decays strengthen the bounds from meson
mixing and, for some regions of the parameter space, even provide the
best limits, see {\it e.g.}  \cite{Silvestrini:2007yf} for a study
with small to moderate $\tan \beta$.

We now consider more model-dependent bounds arising for large
$\tan \beta$, where also the $B \to \ell^+ \ell^-$ decays come into
play.  The dependence of the $\delta_{23}^d$-bounds on $\tan \beta$
and SUSY mass terms can be seen, {\it e.g.}, in \cite{Foster:2006ze}.

An important mechanism for $\tan \beta$ enhancements are Higgs
penguins, magnifying gluino loops with down squark flavor mixing in
$\Delta B=1$, notably $B \to \mu^+ \mu^-$ decays, and $\Delta B=2$
processes, {\it e.g.}, \cite{Isidori:2002qe}.  Using the recent $95
\%$ C.L.~bounds on the branching ratios, ${\cal{B}}(B_d \to \mu^+
\mu^-) <1.8 \cdot 10^{-8}$ and ${\cal{B}}(B_s \to \mu^+ \mu^-) <5.8
\cdot 10^{-8}$ \cite{Harr:2008kj}, we obtain for $\tan \beta =30$,
$x=1$ and $A=L,R$
\beq
 \label{eq:bmumubounds}
|(\delta^d_{13})_A| < 0.04 \cdot \left( \frac{M_{A^0}}{200 \, \mbox{GeV}}
\right)^2 , ~~~~~~
|(\delta^d_{23})_A| < 0.06 \cdot \left( \frac{M_{A^0}}{200 \, \mbox{GeV}}
\right)^2 ,
\eeq
where $M_{A^0}$ denotes the pseudoscalar Higgs mass. The bounds scale
very roughly as $(30/\tan \beta)^3$, and also depend via non-holomorphic
corrections on the higgsino parameters. Since the experimental
limits are a factor of $\sim 10 \, (100)$ away from the corresponding
Standard Model branching ratios for  $B_s (B_d) \to \mu^+ \mu^-$
decays, the bound on $\delta^d_{23}$ is more constraining than the one
on $\delta^d_{13}$.

Bounds in a similar ballpark can be obtained from neutral Higgs
exchange effects in $B_s$ and $B_d$ mixing (for $\tan \beta=30, x=1$)
\cite{Isidori:2002qe}:
\beq \label{eq:bmixbounds}
\langle \delta^d_{13}\rangle  < 0.01 \cdot \left( \frac{M_{A^0}}{200
    \, \mbox{GeV}} \right) , ~~~~~
\langle \delta^d_{23} \rangle < 0.04 \cdot \left( \frac{M_{A^0}}{200
    \, \mbox{GeV}} \right) ,
\eeq
which scale roughly as $(30/\tan \beta)^2$.

The constraints in Eq.~(\ref{eq:bmumubounds}) and
Eq.~(\ref{eq:bmixbounds}) can be stronger than those given in Table
\ref{tab:exp}, but can be evaded by large $M_{A^0}$ and by small $\tan
\beta$. Note that the mixing bounds decouple slower than the $B \to
\mu^+ \mu^-$ ones , so in order to have large effects in the rare
decays, either a very large $\tan \beta$ or a very light Higgs is
required, or a hierarchy between the $(\delta^d_{i3})_L$ and
$(\delta^d_{i3})_R$ parameters such that $\langle \delta^d_{i3}
\rangle$ is small.

%%%%%%%%%%%%%%%%%%%%
\section{Hybrid Gauge-Gravity Mediation
\label{sec:hybrid}}
Ref. \cite{Feng:2007ke} has considered a mediation mechanism that
allows non-MFV contributions to the soft supersymmetry breaking terms,
yet flavor changing terms are naturally suppressed. The basic
assumption is that the gauge-mediated contributions are dominant, but
gravity-mediated contributions are non-negligible. The structure of
the latter is, however, not arbitrary. An approximate Abelian symmetry
which explains the smallness and the hierarchy of the Yukawa couplings
(the Froggatt-Nielsen mechanism) dictates at the same time a flavor
structure for the soft terms.

In this Section, we analyze the predictions of this framework for
the flavor changing $\delta^q_{ij}$ parameters.
We write down the high scale soft terms in Section \ref{sec:GGsoft},
and include effects from renormalization group evolution (RGE) in
Section \ref{sec:flavormz}. Therein we also present the low energy
$\delta^q$ parameters in hybrid mediation. Mass splittings and flavor
mixing matrices are considered in Section \ref{sec:splitmix}.

%%%%%%%%%%%
\subsection{Gauge and gravity soft breaking \label{sec:GGsoft}}
The soft breaking terms for the squarks have then the following
form, at the scale of gauge mediation, $m_M$:
\beqa\label{inicon}
M^2_{\tilde Q_L}(m_M)&=&\tilde m^2_{Q_L} ({\bf 1}+ r X_{Q_L}),\no\\
M^2_{\tilde D_R}(m_M)&=&\tilde m^2_{D_R} ({\bf 1}+r X_{D_R}),\no\\
M^2_{\tilde U_R}(m_M)&=&\tilde m^2_{U_R} ( {\bf 1}+r X_{U_R}),
\eeqa
where $r\lsim1$ parameterizes the ratio between the gravity-mediated
and the gauge-mediated contributions, and is discussed further in
Section \ref{sec:messenger}. While the gauge-mediated
initial conditions are flavor blind, the structure of the $X_{q_A}$ matrices,
coming from gravity mediation, is subject to the selection rules of
the FN symmetry.

The diagonal terms of the $X_{q_A}$ matrices are never suppressed by the
horizontal symmetry. On the other hand, the off-diagonal entries are
suppressed whenever the two corresponding generations carry different
$H$-charges. Within the simplest FN models, with a single horizontal
$U(1)_H$ symmetry, the parametric suppression of the off-diagonal
terms is related to that of the quark parameters:
\beq\label{fnxij}
(X_{q_{L,R}})_{ii}\sim1,\ \ \ (X_{q_L})_{ij}\sim|V_{ij}|,\ \ \
(X_{q_R})_{ij}\sim\frac{m_{q_i}/m_{q_j}}{|V_{ij}|}\ \ \ (i<j), ~~q=U,D.
\eeq
The ``$\sim$'' sign here means ``of the same parametric suppression
as'' but with generally different ${\cal O}(1)$ complex coefficients.

The squark mass-squared matrices $\overline M^2_{\tilde q_A}$
then have the following form:
\beqa
\overline M^2_{\tilde D_L}&=&M ^2_{\tilde Q_L}+D_{D_L}{\bf 1}
+m_Dm_D^\dagger ,\no\\
\overline M^2_{\tilde U_L}&=&M^2_{\tilde Q_L}+D_{U_L}{\bf 1}
+m_Um_U^\dagger ,\no\\
\overline M^2_{\tilde D_R}&=&M^2_{\tilde D_R}+D_{D_R}{\bf 1}
+m_D^\dagger m_D ,\no\\
\overline M^2_{\tilde U_R}&=&M^2_{\tilde U_R}+D_{U_R}{\bf 1}
+m_U^\dagger m_U,
 \label{eq:squark-masses}
\eeqa
where $m_{U,D}$ are the up and down quark mass matrices in the flavor basis,
$D_{q_A}$ are the $D$-term contributions and all quantities
should be evaluated at the electroweak scale
$\mu \sim m_Z$. We assume that $r>y_t^2|V_{ts}|^2\sim0.002$, so that
the gravity-mediated contributions are non-negligible.

%%%%%%%%%%%
\subsection{Flavor breaking at $m_Z$ \label{sec:flavormz}}
The initial conditions (\ref{inicon}) hold at the scale of
gauge mediation, $m_M$, and the flavor relations (\ref{fnxij})
hold at the scale of gravity mediation, the Planck mass $m_{\rm Pl}$.
We are, however, interested in the predictions for the
$(\delta^{q}_{ij})_A$
parameters, requiring soft terms evaluated at the electroweak scale.
We thus need to take into account the effects of renormalization
group evolution. A detailed discussion of the RGE is given
in Appendix \ref{app:rge}.  The final conclusions are the following:

(i) Starting from the soft squark masses at the scale $m_M$
of the form given in Eq. (\ref{inicon}), the soft squark masses at
the scale $m_Z$ can be written in the following approximate form:
\beqa\label{eq:msoftmz}
M^2_{\tilde Q_L}(m_Z)& \sim&\tilde m^2_{Q_L} (r_3 {\bf 1}+
c_u Y_u Y_u^\dagger + c_d Y_d Y_d^\dagger+ r X_{Q_L}), \nonumber\\
M^2_{\tilde U_R}(m_Z)&\sim&\tilde m^2_{U_R} (r_3 {\bf 1}+
c_{uR} Y_u^\dagger Y_u+ r X_{U_R}), \nonumber \\
M^2_{\tilde D_R}(m_Z)& \sim&\tilde m^2_{D_R} (r_3 {\bf 1}+
c_{dR} Y_d^\dagger Y_d + r X_{D_R}),
\eeqa
where $Y_u$ and $Y_d$ denote the up and down quark Yukawa matrices
in the  flavor basis.

(ii) The relations between the off-diagonal elements $(X_{q_{L,R}})_{ij}$
and the quark parameters, given in Eq.~(\ref{fnxij}), are either
RGE-invariant to a good approximation, or changed by factors of ${\cal
   O}(1)$. In any case, the relations between the parametric
suppressions remain the same, and one should simply use the low
energy values of $|V_{ij}|$ and of $m_{q_i}/m_{q_j}$ to estimate
the low energy values of $(X_{q_{L,R}})_{ij}$.

(iii) We define the factor $r_3$ via the RGE correction to the
diagonal elements of the soft squark mass matrices
$(M^2_{\tilde q_A})_{ii}$:
\beq\label{defrthree}
\tilde m^2_{12}(\mu=m_Z)=r_3\tilde m^2_{12}(\mu=m_M) ,
\eeq
with the average diagonal mass-squared defined as
\beq \label{eq:m12tilde}
\tilde m^2_{ij} \equiv \frac{1}{2}
\left( (M^2_{\tilde q_A})_{ii}+(M^2_{\tilde q_A})_{jj} \right) .
\eeq
In writing Eqs. (\ref{defrthree}) and (\ref{eq:m12tilde}) with
the same $\tilde m_{12}^2$ and $r_3$ for all three sectors ($\tilde Q_L,
\tilde U_R,\tilde D_R$) we take into account that the dominant
contribution to the initial squark soft masses and to their RGE is
QCD-induced and, in the limit that we neglect the electroweak gauge
couplings, is universal among all squarks.
Numerically, $r_3$ is of ${\cal{O}}(1-10)$, depending on the initial
conditions and the scale of supersymmetry breaking.  Details on $r_3$
in gauge mediation are given in Section \ref{sec:messenger}.
In minimal models, typically $r_3 \sim 3$.

(iv) The coefficients $c_u,c_d,c_{uR},c_{dR}$ are of order
$[5 /(16 \pi^2)] \ln (m_M/m_Z)$ and can be ${\cal{O}}(1)$ for
$m_M \sim m_{\rm GUT}$ (see, {\it e.g.} Ref. \cite{Paradisi:2008qh}
for numerical formulae). All coefficients $c_u,c_d,c_{uR},c_{dR}<0$.
Hence, the Yukawa corrections reduce the low energy values of the diagonal
$(M^2_{\tilde q_A})_{33}$ entries with respect to the high energy ones.
Note that we neglect subdominant (MFV) terms with higher powers of the Yukawa
couplings; the general form of the MFV soft terms is given in Ref.
\cite{D'Ambrosio:2002ex}.

Before we derive our order of magnitude estimates for the various
$\delta^q_{ij}$ parameters, two comments are in order:
\begin{enumerate}
\item
In the following we use the various $\tilde m_{ij}^2(m_Z)$ to evaluate
the denominator of the $(\delta^q_{ij})_A$ parameters instead of using
the physical mass average as in Section \ref{sec:fcnc}.  In this way
we neglect $D$-terms of ${\cal{O}}(m_Z^2/\tilde m_{ij}^2)$ and $F$-terms
of at most ${\cal{O}}(m_t^2/\tilde m_{i3}^2)$.  It is straightforward
to include such corrections into our analysis, but since the flavor
pattern from FN gravity is only accurate up to order one numbers,
this does not improve the precision of our predictions.
\item Eq.~(\ref{eq:msoftmz}) is written in the
flavor basis. We can read off the $\delta^q$ parameters
after rotating the squarks by the same transformation that brings the quarks
to mass eigenstates, see Eq.~(\ref{eq:deltaq-def}). This rotation does not
change the parametric suppression of the $X_{ij}$ terms, and therefore we can
still use the estimates (\ref{fnxij}) in the new basis. The rotation
can affect the order one coefficients in these terms, but these are
unknown anyway.
\end{enumerate}

We now write the low energy values of the entries in the squark
mass matrices in the basis where the quark mass
matrices and gluino couplings are diagonal. We are interested in models with
$r>y_t^2|V_{ts}|^2$, in which case the gravity-mediated contributions are
non-negligible (see below). We can thus neglect all Yukawa couplings except
third generation ones. For MFV contributions, we use notations such as $V_{td}$ to denote
the actual contributing CKM element. For the non-MFV contributions,
where there is uncertainty of order one, we use, for example, the
notation $V_{13}$ to represent parametric suppression that is similar
to that of $V_{ub}$ or $V_{td}$.  We obtain ($q=U,D$, $i \neq 3$):
\beqa
(\widetilde M^2_{\tilde q_L}(m_Z))_{33}& \sim &
\tilde m^2_{Q_L} (r_3 + c_u y_t^2 + c_d y_b^2+r), \nonumber \\
(\widetilde M^2_{\tilde q_L}(m_Z))_{ii}& \sim &
\tilde m^2_{Q_L} (r_3 +r) , \nonumber \\
(\widetilde M^2_{\tilde U_L}(m_Z))_{12}& \sim &
\tilde m^2_{Q_L} (c_d y_b^2 V_{ub} V_{cb}^*+r |V_{12}|) ,\nonumber \\
(\widetilde M^2_{\tilde U_L}(m_Z))_{i3}&\sim &
\tilde m^2_{Q_L} (c_d y_b^2 V_{ib} V_{tb}^*+r |V_{i3}|) ,\nonumber \\
(\widetilde M^2_{\tilde D_L}(m_Z))_{12}& \sim &
\tilde m^2_{Q_L} (c_u y_t^2 V_{ts} V_{td}^*+r |V_{12}|) ,\nonumber \\
(\widetilde M^2_{\tilde D_L}(m_Z))_{i3}& \sim &
\tilde m^2_{Q_L} (c_u y_t^2 V_{tb} V_{ti}^*+r |V_{i3}|) .
\label{eq:mql2-soft}
\eeqa
Hence, with $r \ll r_3$,
\beqa\label{eq:delql}
(\delta^{u}_{12})_L & \sim & \frac{|V_{12}|}{r_3}
{\rm max}(r, c_d y_b^2|V_{ub} V_{cb}^*/V_{12}|)
\sim r \frac{|V_{12}|}{r_3} ,\no \\
(\delta^{d}_{12})_L & \sim & \frac{|V_{12}|}{r_3}
{\rm max}(r, c_u y_t^2|V_{ts} V_{td}^*/V_{12}|)\sim
r \frac{|V_{12}|}{r_3} ,\no \\
(\delta^{u}_{i3})_L& \sim & \frac{|V_{i3}|}{r_3}
{\rm max} (r, c_d y_b^2)
\sim \hat r \frac{|V_{i3}|}{r_3} ,\no\\
(\delta^{d}_{i3})_L& \sim & \frac{|V_{i3}|}{r_3}
{\rm max} (r, c_u y_t^2)
\sim \frac{|V_{i3}|}{r_3} ,\no\\
\delta^{cu}_{i3} & \simeq & (\delta^{d}_{i3})_L ,
\eeqa
where
\beq\label{eq:rhat}
\hat r\equiv{\rm max}\{r,y_b^2\}.
\eeq
Given that $y_b^2\sim0.001\tan^2\beta$, the distinction between $\hat r$
and $r$ is important only if $\tan\beta$ is large.

We can now explain our choice to focus on the region of $r>y_t^2|V_{ts}|^2$.
If $r$ were smaller than that, then MFV contributions would dominate
$(\delta^{d}_{12})_L$ and, for $\tan\beta\gsim10$, also  $(\delta^{u}_{12})_L$.

For the $(\delta^q_{ij})_R$, $q=U,D$ we obtain $i \neq 3$, $j=1,2,3$:
\beqa
(\widetilde M^2_{\tilde U_R}(m_Z))_{33}& \sim &
\tilde m^2_{U_R} (r_3 + c_{uR} y_t^2 +r), \nonumber \\
(\widetilde M^2_{\tilde D_R}(m_Z))_{33}& \sim &
\tilde m^2_{D_R} (r_3 + c_{dR} y_b^2 +r), \nonumber \\
(\widetilde M^2_{\tilde q_R}(m_Z))_{ii}& \sim &
\tilde m^2_{q_R} (r_3 +r) , \nonumber \\
(\widetilde M^2_{\tilde q_R}(m_Z))_{ij}& \sim&
\tilde m^2_{q_R} r \frac{m_{q_i}}{m_{q_j} |V_{ij}|} , \label{eq:mqr2-soft}
\eeqa
hence
\beqa\label{eq:delqr}
(\delta^{q}_{ij})_R & \sim & \frac{r}{r_3}
\frac{m_{q_i}}{m_{q_j} |V_{ij}|} .
\eeqa

\begin{table}[t]
\caption{The order of magnitude estimates for
  $(\delta_{ij}^{d,u})_{L,R}$ and $\langle\delta^{d,u}_{ij}\rangle$ in
  the hybrid gauge-gravity models. The numerical estimates are
  obtained using quark masses at the scale $m_Z$ \cite{Xing:2007fb},
  and taking $r_3=3$. All results scale as $(3/r_3)$. }
\label{tab:the}
\begin{center}
\begin{tabular}{ll|ccc} \hline\hline
\rule{0pt}{1.2em}%
$q$ & $ij$\ &  $(\delta^q_{ij})_{L}$ &  $(\delta^q_{ij})_{R}$ &
$\langle\delta^q_{ij}\rangle$ \cr \hline
$d$ & $12$\ & $(r/r_3)|V_{12}|\sim0.08r$ & $\frac{(r/r_3)(m_d/m_s)}{|V_{12}|}\sim0.08r$
& $(r/r_3)\sqrt{m_d/m_s}\sim0.08r$ \cr
$d$ & $13$\ & $|V_{13}|/r_3\sim0.001$ &
$\frac{(r/r_3)(m_d/m_b)}{|V_{13}|}\sim0.08 r$
& $\sqrt{r m_d/m_b}/r_3\sim0.01\sqrt{r}$ \cr
$d$ & $23$\ & $|V_{23}|/r_3\sim0.01$ &
$\frac{( r/r_3)(m_s/m_b)}{|V_{23}|}\sim0.2 r$
& $\sqrt{r m_s/m_b}/r_3\sim0.05\sqrt{ r}$ \cr
$u$ & $12$\ & $(r/r_3)|V_{12}|\sim0.08r$ & $\frac{(r/r_3)(m_u/m_c)}{|V_{12}|}\sim0.003r$
& $(r/r_3)\sqrt{m_u/m_c}\sim0.02r$  \cr
$u$ & $13$\ & $(\hat r/r_3)|V_{13}|\sim0.001 \hat r$ & $\frac{(r/r_3)(m_u/m_t)}{|V_{13}|}\sim0.0006 r$
& $\sqrt{r \hat r m_u/m_t}/r_3\sim0.0009\sqrt{ r \hat r}$ \cr
$u$ & $23$\ & $(\hat r/r_3)|V_{23}|\sim0.01 \hat r$ & $\frac{(r/r_3)(m_c/m_t)}{|V_{23}|}\sim0.03 r$
& $\sqrt{r \hat r m_c/m_t}/r_3\sim0.02\sqrt{ r \hat r}$ \cr
\hline\hline
\end{tabular}
\end{center}
\end{table}

We finally obtain the order of magnitude estimates for
the $\delta_{ij}^q$ parameters presented in Table \ref{tab:the}.
We would like to emphasize the following points:
\begin{enumerate}
\item The RGE suppresses the flavor violating
$\delta^q$ parameters.
\item The values of $(\delta^d_{i3})_L$ are independent of $r$.
The reason for this are the RGE-induced ${\cal{O}}(y_t^2)$ terms which
dominate the gravity-mediated ones of order $r$.

\item The values of $\langle\delta^q_{ij}\rangle$ are independent of the
   CKM parameters.
\end{enumerate}

One of the issues that we are trying to clarify is whether one
can differentiate between MFV and non-MFV mediation of supersymmetry
breaking. Indeed, our framework gives contributions to
$(\delta^q_{ij})_R$ that cannot be achieved in MFV models.
The parameters $(\delta^d_{ij})_L$, however, receive a contribution
from MFV initial conditions (such as pure gauge mediation), which is
CKM induced and of the order
$(V_{tj} V_{ti}^*/r_3) [y_t^2/(16 \pi^2)] \ln(m_{M}/m_Z)$
times a numerical factor of ${\cal{O}}(5)$ (see Appendix
\ref{app:rge}). For $j=3$ this is the dominant contribution and,
therefore, $(\delta^d_{i3})_L$ itself is not indicative of hybrid
mediation. For $r<y_t^2|V_{ts}|^2$, even the $(\delta^d_{12})_L$
would be dominated by the MFV contribution.
A similar comment applies to $(\delta^u_{ij})_L$ for
large $\tan \beta$ due to the $V_{ib} V_{jb}^* y_b^2$ induced
RGE contribution.

%%%%%%%%%%%
\subsection{Splittings and mixing \label{sec:splitmix}}
A flavor changing $\delta_{ij}$ parameter depends on three factors:
the overall squark mass scale $\tilde m_{ij}$, the mass splitting
$\Delta\tilde m^2_{ij}$, and the mixing angle $K_{ij}$. While low
energy measurements of FCNC
processes are sensitive only to the $\delta^q_{ij}$ parameters,
high-$p_T$ experiments can, in principle, measure each of these
three ingredients separately, hence providing further information
regarding the supersymmetric flavor structure \cite{Feng:2007ke}. It is thus
of interest to estimate $\Delta\tilde m^2_{ij}/\tilde m_{ij}^2$ and
$K_{ij}$ in our hybrid gauge-gravity framework.

Investigation of Eqs.~(\ref{eq:mql2-soft}), (\ref{eq:mqr2-soft}) and
the analysis of Appendix  \ref{app:rge} leads to the following
estimates of the $m_Z$-scale mass splittings:
\beqa\label{msplit}
\frac{\Delta\tilde m^2_{12}}{\tilde m^2_{12}}&\sim&
\begin{array}{cc} r/r_3 & \ \ \
   (\tilde D_L,\tilde U_L,\tilde D_R,\tilde U_R) \end{array} , \cr
\frac{\Delta\tilde m^2_{i3}}{\tilde m^2_{i3}}&\sim&
\left\{\begin{array}{cc}
     1/r_3 & (\tilde D_L,\tilde U_L,\tilde U_R) \cr
     \hat r/r_3 & (\tilde D_R)
   \end{array}\right. ~~~~~~~\mbox{for}~~ i \neq 3 .
\eeqa

As concerns the mixing matrices, they depend on the unitary matrices
that diagonalize the various quark and squark mass matrices. We define:
\beqa
V^d_L m_D V^{d\dagger}_R&=&{\rm diag}(m_d,m_s,m_b),\cr
V^u_L m_U V^{u\dagger}_R&=&{\rm diag}(m_u,m_c,m_t),\cr
\tilde V^d_A \overline M^2_{\tilde D_A}\tilde V^{d\dagger}_A&=&
{\rm diag}(\tilde m^2_{\tilde d_{A1}},\tilde m^2_{\tilde d_{A2}},
\tilde m^2_{\tilde d_{A3}}),\cr
\tilde V^u_A \overline M^2_{\tilde U_A}\tilde V^{u\dagger}_A&=&
{\rm diag}(\tilde m^2_{\tilde u_{A1}},\tilde m^2_{\tilde u_{A2}},
\tilde m^2_{\tilde u_{A3}}),
\eeqa
where $A=L,R$. We obtain for the mixing matrices relevant in neutral
gaugino couplings
\beq
K^q_A=V^q_A \tilde V^{q\dagger}_A,
\eeq
and for the quark mixing matrix:
\beq
V=V^u_L V_L^{d \dagger} .
\eeq

The parametric suppression of the off-diagonal terms in $V^q_A$
in the FN basis (that is, the basis where the FN charges are
well-defined) is determined by the quark flavor parameters:
\beqa\label{eq:vfn}
(V^d_L)_{ij}&\sim&|V_{ij}|,\no\\
(V^u_L)_{ij}&\sim&|V_{ij}|,\no\\
(V^d_R)_{ij}&\sim&\frac{m_{d_i}/m_{d_j}}{|V_{ij}|},\no\\
(V^u_R)_{ij}&\sim&\frac{m_{u_i}/m_{u_j}}{|V_{ij}|}.
\eeqa
The parametric suppression of the off-diagonal terms in $\tilde V^q_A$
in the FN basis is determined by $r$ and by the quark flavor parameters:
\beq\label{eq:susyk}
\begin{array}{ccc}
(\tilde V^d_L)_{12}\sim|V_{12}|, & &
(\tilde V^d_L)_{i3}=(V^u_L)_{i3}+{\cal O}(\hat r|V_{i3}|),\cr
(\tilde V^u_L)_{12}\sim|V_{12}|, & &
(\tilde V^u_L)_{i3}=(V^u_L)_{i3}+{\cal O}(\hat r|V_{i3}|),\cr
(\tilde V^d_R)_{12}\sim\frac{m_d/m_s}{|V_{12}|}, & &
(\tilde V^d_R)_{i3}=(V^d_R)_{i3}+
{\cal O}(\frac{r(m_{d_i}/m_b)}{\hat r|V_{i3}|}),\cr
(\tilde V^u_R)_{12}\sim\frac{m_u/m_c}{|V_{12}|}, & &
(\tilde V^u_R)_{i3}=(V^u_R)_{i3}+{\cal O}(
\frac{r(m_{u_i}/m_t)}{|V_{i3}|}).\end{array}
\eeq
We note the following points, which can be further understood on the
basis of our analysis in Appendix \ref{app:rge}:
\begin{enumerate}
\item In the up quark mass basis, $(\tilde V^d_L)_{i3}\sim (\tilde
V^u_L)_{i3}\sim \hat r|V_{i3}|$. The reason is that in this basis the $Y_u
Y_u^\dagger$ term in the RGE is diagonal, and the leading non-diagonal
contribution is either the $r$-suppressed gravity-mediated contribution
or the $y_b^2$-suppressed MFV contribution.
\item In the up quark mass basis, $(\tilde V^u_R)_{i3}\sim
r(m_{u_i}/m_t)/|V_{i3}|$. The reason is that in this basis the $Y_u^\dagger Y_u$ term in the RGE is diagonal, and the leading non-diagonal
contribution is the $r$-suppressed gravity-mediated contribution.
\item In the down quark mass basis, $(\tilde V^d_R)_{i3}\sim
(r/\hat r)(m_{d_i}/m_b)/|V_{i3}|$. The reason is that in this basis the $Y_d^\dagger Y_d$ term in the RGE is diagonal, and the leading non-diagonal
contribution is the $r$-suppressed gravity-mediated contribution.
\end{enumerate}

We thus find
\beqa\label{K1}
(K_L^d)_{12}&\sim& |V_{12}|,\ \ \ (K_L^d)_{i3}\sim |V_{ti}|,\no\\
(K_L^u)_{12}&\sim&|V_{12}|,\ \ \ (K_L^u)_{i3}\sim\hat r|V_{i3}|,\no\\
(K_R^d)_{12}&\sim&\frac{m_d/m_s}{|V_{12}|},\ \ \
(K_R^d)_{i3}\sim\frac{r(m_{d_i}/m_b)}{\hat r|V_{i3}|},\no\\
(K_R^u)_{12}&\sim&\frac{m_u/m_c}{|V_{12}|},\ \ \
(K_R^u)_{i3}\sim\frac{r(m_{u_i}/m_t)}{|V_{i3}|}.
\eeqa

%%%%%%%%%%%%%%%%%%%%%
\section{Phenomenological consequences}
\label{sec:phe}
By comparing the phenomenological constraints of Table \ref{tab:exp}
to the theoretical order of magnitude predictions of the hybrid
gauge-gravity models of Table \ref{tab:the}, we can put an upper bound
on $r$ and on $\hat r$, and describe the possible FCNC effects of the model.
The strongest bound on $r$ comes from the $\langle\delta^d_{12}\rangle$
parameter, and it reads
\beq
  \label{eq:roverr3}
r/r_3\lsim0.01-0.03.
\eeq
We use here $m_{\tilde q}=1$ TeV; the bounds would be stronger by
$m_{\tilde q}/(1$ TeV) for lighter $m_{\tilde q}$. The stronger bound
corresponds to $x=1$ and a phase of order $0.3$, while the weaker
bound corresponds to $x=4$ and a phase smaller than $0.1$.
The $\hat r$ parameter affects only the $\delta^u_{i3}$ parameters,
so there is no phenomenological constraint on its size, and it
is only bounded by its definition:

\beq
 \label{eq:rhatoverr3}
r\leq \hat r \lsim 1.
\eeq
For small values of $\tan \beta$, $\hat r =r$ and Eq.~(\ref{eq:roverr3})
applies to $\hat r$.  Inserting $r/r_3\lsim0.03$ and $r\leq\hat r\lsim1$
into the predictions of Table \ref{tab:the}, we obtain the upper
bounds on the $\delta^q_{ij}$ given in Table \ref{tab:upb}.

\begin{table}[t]
\caption{The order of magnitude upper bounds on
   $(\delta_{ij}^{d,u})_{L,R}$ and
   $\langle\delta^{d,u}_{ij}\rangle$ for $r/r_3\lsim0.03$.
   Entries in parenthesis are independent of
   $r$, therefore representing estimates rather than upper bounds, and
   scale as $(3/r_3)$. The bounds on $\langle \delta^d_{13,23}\rangle$ scale as $\sqrt{3/r_3}$.  The bounds on $(\delta^u_{i3})_L$
   [$\langle\delta^u_{i3}\rangle$] correspond to $\hat r\sim1$ and
   scale as $(3/r_3)$ [$\sqrt{3/r_3}$]; if $\hat r=r$, these bounds
    are a factor of 10 [$\sqrt{10}$] stronger and do not scale
   with $r_3$.     }
\label{tab:upb}
\begin{center}
\begin{tabular}{ll|ccc} \hline\hline
\rule{0pt}{1.2em}%
$q$ & $ij$\ &  $(\delta^q_{ij})_{L}$ &  $(\delta^q_{ij})_{R}$ &
$\langle\delta^q_{ij}\rangle$ \cr \hline
$d$ & $12$\ & $0.007$ & $0.007$ & $0.007$ \cr
$d$ & $13$\ & $[0.001]$ & $0.007$ & $0.003$ \cr
$d$ & $23$\ & $[0.01]$ & $0.01$ & $0.01$ \cr
$u$ & $12$\ & $0.007$ & $0.0003$ & $0.001$  \cr
$u$ & $13$\ & $0.001$ & $0.00005$ & $0.0003$ \cr
$u$ & $23$\ & $0.01$ & $0.003$ & $0.006$ \cr
\hline\hline
\end{tabular}
\end{center}
\end{table}

\begin{figure}
\includegraphics[scale=1.0]{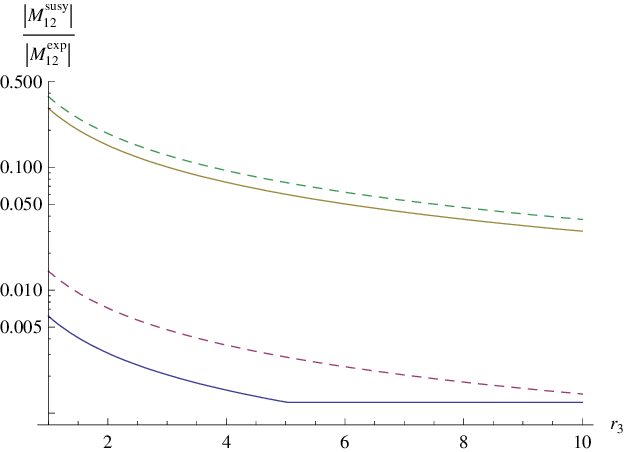}
\caption{\label{fig:range} Maximum reach in $B_d$ (solid) and
$B_s$ (dashed) mixing, $|M_{12}^{\rm susy}/M_{12}^{\rm exp}|$,
as a function of the RGE-factor $r_3$. The uppermost two
curves correspond to $\tan \beta=30$ and
$M_{A^0}=200 \, \mbox{GeV}$.}
\end{figure}

We then learn that the maximal possible effects in the neutral $B_d$,
$B_s$ and $D$ systems, are as follows (for $r_3=3$):
\beq\label{eq:FN-reach}
\begin{array}{lc}
B_d: & |M_{12}^{\rm susy}/M_{12}^{\rm exp}|\lsim0.002,\\
B_s: & |M_{12}^{\rm susy}/M_{12}^{\rm exp}|\lsim0.005,\\
D: & |M_{12}^{\rm susy}/M_{12}^{\rm exp}|\lsim0.05. \end{array}
\eeq
Note that for $D$-meson mixing, we use for $M_{12}^{\rm exp}$ the
experimental upper bound. The stronger this bound will become, the
more significant role the SUSY contribution can play.

We emphasize the following points:
\begin{enumerate}
\item The bound in the $D$ system comes from
$\langle\delta^u_{12}\rangle$ and is $r_3$ independent.
\item For $r_3={\cal O}(1-10)$, the bound in the $B_s$
system comes from $\langle\delta^d_{23}\rangle$ and scales as
 $3/r_3$.
\item For $r_3={\cal O}(1-5)$, the bound in the $B_d$
system comes from $\langle\delta^d_{13}\rangle$ and scales as
 $3/r_3$. For $r_3>5$, the bound comes from $(\delta^d_{13})_R$
 and does not scale with $r_3$.
\end{enumerate}

For large $\tan \beta$ and low $M_{A^0}$, the $B_{d,s}$ mixing
amplitudes can be significantly enhanced, as discussed in Section
\ref{sec:delF1}. Comparing the phenomenological constraints of
Eq.~(\ref{eq:bmixbounds}) to Table \ref{tab:the}, we obtain for
$r_3=3$ (and $\tan \beta=30$, $M_{A^0}=200 \, \mbox{GeV}$):
\beq\label{eq:FN-reach-largetb}
\begin{array}{lc}
B_d: & |M_{12}^{\rm susy}/M_{12}^{\rm exp}|\lsim0.10,\\
B_s: & |M_{12}^{\rm susy}/M_{12}^{\rm exp}|\lsim0.13 .
\end{array}
\eeq
The $r_3$ dependence of upper bounds on the supersymmetric
contributions to $B_d$ and $B_s$ mixings is shown in Fig.~\ref{fig:range}.

We now discuss where further signals of this non-MFV scenario could
arise.  While the bounds from Eq.~(\ref{eq:bmumubounds}) and
(\ref{eq:bmixbounds}) can be evaded for suitable values of $M_{A^0}$
and $\tan \beta$, they indicate on the other hand that in FN gravity
models an observation of $B_{d,s} \to \mu^+ \mu^-$ decays is possible
near their current experimental limits.  This
itself is, however, not a unique sign of our model, since it can
happen also in the MFV MSSM at large $\tan \beta$, e.g.,
\cite{Buras:2002vd}.  One crucial difference is the breakdown of MFV
relations between $b \to s$ and $b \to d$, such as in $B \to \mu^+
\mu^-$ decays \cite{Bobeth:2002ch}.  We find for the ratio $R_{\mu
  \mu}$
\beqa \label{eq:Rmumu}
R_{\mu \mu}=\frac{{\cal{B}}(B_s \to \mu^+ \mu^-)}{{\cal{B}}(B_d \to
  \mu^+ \mu^-)} \sim
\frac{m_{B_s} f_{B_s}^2 \tau_{B_s}}{m_{B_d} f_{B_d}^2 \tau_{B_d}}
\times r_{\rm ps} \times
\left\{\begin{array}{cl}
  \frac{|V_{ts}|^2 }{|V_{td}|^2} & \mbox{for}~~ (\mbox{MFV},
  (\delta^d_{i3})_L) ,\cr
   \frac{|m_s V_{td}|^2}{|m_d V_{ts}|^2} & \mbox{for}~~
   ((\delta^d_{i3})_R) , \cr
 \frac{m_s}{m_d} &  \mbox{for}~~ (\langle \delta^d_{i3} \rangle) ,
   \end{array}\right.
\eeqa
where $f_{B_q}$, $m_{B_q}$ and $\tau_{B_q}$
denote the decay constant, mass and lifetime of the
$B_q$, $q=d,s$, respectively, and $r_{\rm ps}$  collects all further, small
(known) U-spin breaking of $R_{\mu \mu}$ related to kinematical factors.

While stemming from qualitatively very different expressions,
numerically the three ratios in Eq.~(\ref{eq:Rmumu}) turn out to be
similar, that is (from top to bottom), 25, 14 and 19, using central
values at $m_Z$ from \cite{Xing:2007fb}.  Since we cannot distinguish
the case with dominant $(\delta^d_{i3})_L$ from MFV, some contribution
from $(\delta^d_{i3})_R$ is required to identify non-MFV. If this is
the case, $R_{\mu\mu}$ is suppressed w.r.t.~its MFV (and Standard
Model) value.
Since there is no large hierarchy between $R_{\mu \mu}$ in the different
scenarios, establishing the FN flavor quantum numbers in this observable
needs a  measurement at the ${\cal{O}}(10\%)$ level
$(3 \sigma)$ and very good control over $f_{B_s}/f_{B_d}$.

We close with some general comments.  Signals of a FN gravity
contribution are those of non-MFV models, that is, {\it e.g.},
\cite{Hiller:2003di},
({\it i}) beyond CKM CP-violation,
({\it ii}) wrong chirality contributions to FCNCs,
and
({\it iii}) the breakdown of CKM-relations as in $R_{\mu \mu}$.
Because the FN gravity model contains only a controlled amount of flavor
violation, an experimental verification needs precise measurements.

Since in FN gravity $(\delta^d_{i3})_R \gsim (\delta^d_{i3})_L$, see
Table \ref{tab:upb}, the natural place to look for such contributions
is in right-handed currents. The sensitivity will be even higher if
one looks in addition for CP-violation. Potentially interesting here
are CP asymmetries in $B \to K^* ( \to K \pi) \ell^+ \ell^-$ decays
\cite{Bobeth:2008ij}.

The impact of charged wino loops to $b$-physics observables is limited
by $(\alpha_2/\alpha_3)$ with respect to the impact of
$(\delta^d_{i3})_L$, see Eq.~(\ref{eq:delql}), and is hence
sub-dominant. Charged higgsino effects could be of interest at large
$\tan \beta$.  Further study is needed.

Note that there is also the possibility of a light stop
having a macroscopic lifetime of order picoseconds,
if the FCNC decay of $\tilde t_1$ to charm plus the lightest
neutralino induced by $\delta^u_{23}$ is sufficiently suppressed yet is the
dominant decay mode \cite{Hiller:2008wp}. The latter can be arranged
kinematically by a small mass splitting, $\Delta M$, between
the $\tilde t_1$ and the lightest neutralino.
In FN gravity a long-lived stop requires the lightest stop to be predominantly
left-handed and $\tan \beta$ to be small, such that
$(\delta^u_{23})_L \lsim 10^{-6} \, (m_{\tilde t_1}/\Delta M)$.
This gives an upper bound $r/r_3 \lsim 3 \cdot 10^{-4}$ for
$\Delta M/m_{\tilde t_1} =0.1$, stronger than the one in
Eq.~(\ref{eq:roverr3}).

%%%%%%%%%%%%%%%%%%%%%%%%%
\section{Holomorphic zeros}
\label{sec:hol}
With a more complicated model employing the Froggatt-Nielsen
mechanism, one can suppress the supersymmetric mixing angles compared
to the values given in Eqs.~(\ref{eq:vfn})~and~(\ref{eq:susyk}), while keeping the
parametric suppression of the quark masses and of the CKM angles
consistent with the measured values \cite{Nir:1993mx}. The horizontal
symmetry has to be extended to, for example, $U(1)_1\times U(1)_2$,
and holomorphic zeros must play a role. At least one of the two
horizontal $U(1)$'s is broken by a single spurion, and some of the
Yukawa couplings carry charge of the same sign as the spurion, and
thus are forbidden by holomorphy.

Originally, this mechanism was used to obtain phenomenologically
viable models without any squark degeneracy. However, recent
improvements in the bound on the mass splitting in the neutral $D$
system imply that degeneracy between the first two generations of
squark doublets at the level of ${\cal O}(10\%)$ or stronger is
required (for squarks lighter than TeV)
\cite{Ciuchini:2007cw,Nir:2007ac,Feng:2007ke}.

Thus, in this section, we investigate the possibility of constructing
such FN-type models, where the required minimal degeneracy comes from
either the gauge-mediation dominance or RGE or both. In particular, we
ask what are the maximal possible effects in the neutral $D,B_d$ and
$B_s$ systems in such a framework.

It was proven in Ref. \cite{Nir:2002ah} that, to obtain
\beq
(K^d_L)_{12}\ll|V_{12}|,\ \ \ (K^d_R)_{12}\ll\frac{m_d/m_s}{|V_{12}|},
\eeq
(as necessary to relax the strong degeneracy requirement), while
keeping the CKM elements large enough, there should be four (and only
four) specific holomorphic zeros in the down quark mass matrix,
leading to both lower and upper bounds on the supersymmetric mixing
angles. These bounds are given in Table \ref{tab:uplow}. The parameter
$\epsilon_{\rm max}$ stands for the largest among the spurions that
break the horizontal FN symmetry. As before, for MFV contributions
(namely those that survive in the $r=0$ limit) we use the notation
$V_{ti}$ rather than $V_{i3}$. The $(K^d_L)_{i3}$ angles get
comparable contributions from MFV and non-MFV sources, so we use the
$V_{i3}$ notations for these.

\begin{table}[t]
\caption{Bounds on the supersymmetric mixing angles in models of
  alignment with suppressed $(K^d_{L,R})_{12}$. For the numerical
  estimates we use quark masses at the scale $m_Z$ \cite{Xing:2007fb}
  and take $r\leq\hat r\lsim1$, and $\epsilon_{\rm max}\sim0.2$. }
\label{tab:uplow}
\begin{center}
\begin{tabular}{l|cc} \hline\hline
\rule{0pt}{1.2em}%
Mixing angle & Lower bound &  Upper bound \cr \hline
$(K^d_L)_{12}$ & $|V_{td}V_{ts}|/r\sim0.0005/r$ &
$|V_{12}|\epsilon_{\rm max}^2\sim0.009$ \cr
$(K^d_R)_{12}$ & $\frac{m_d}{m_s}|V_{13}V_{23}|\sim9\cdot10^{-6}$ &
$\frac{m_d}{m_s |V_{12}|}\epsilon_{\rm max}^2\sim0.009$ \cr
$(K^d_L)_{13}$ & $|V_{13}|\sim0.004$ & $|V_{13}|\sim0.004$ \cr
$(K^d_R)_{13}$ & $\frac{m_d}{m_b}|V_{13}|\sim4\cdot10^{-6}$ &
$\frac{r}{\hat r}\frac{m_d}{m_b |V_{13}|}\epsilon_{\rm max}^2\sim0.009$ \cr
$(K^d_L)_{23}$ & $|V_{23}|\sim0.04$ & $|V_{23}|\sim0.04$ \cr
$(K^d_R)_{23}$ & $\frac{m_s}{m_b}|V_{23}|\sim0.0008$ &
$\frac{r}{\hat r}\frac{m_s}{m_b |V_{23}|}\epsilon_{\rm max}^2\sim0.02$ \cr
$(K^u_L)_{12}$ & $|V_{12}|\sim0.2$ & $|V_{12}|\sim0.2$ \cr
$(K^u_R)_{12}$ & $\frac{m_u}{m_c}|V_{12}|\sim0.0005$ &
$\frac{m_u}{m_c |V_{12}|}\sim0.009$ \cr
\hline\hline
\end{tabular}
\end{center}
\end{table}

The analysis of the $(K^d_L)_{12}$ requires some explanation.
The $\tilde d_L-\tilde s_L$ block of $\widetilde M^2_{\tilde D_L}(m_Z)$
has the following form:
\beq
\widetilde M^2_{\tilde D_L}(m_Z)\sim\tilde m^2_{D_L}
\left( \begin{array}{cc}
r_3 +rX_{11} & c_u y_t^2 V_{td}^*V_{ts}+rX_{12} \\
 c_u y_t^2 V_{td}V_{ts}^*+rX_{12}^* & r_3+rX_{22}+ c_u y_t^2|V_{ts}|^2
\end{array}\right).
\eeq
Here $X_{11}$ and $X_{22}$ are ${\cal O}(1)$ and different from
each other, while $X_{12}$ is taken to lie in the range
$(0,|V_{12}|\epsilon_{\rm max}^2)$. We remind the reader that
we restrict our analysis to the region where $r$ is larger than
$y_t^2|V_{ts}|^2$, so the latter term can be neglected in the
(2,2) entry. The lower bound on $(K^d_L)_{12}$ corresponds to
a negligibly small $X_{12}$. The upper bound given in
the table corresponds to $|X_{12}|\sim|V_{12}|\epsilon_{\rm max}^2$
and $r\gsim0.05$. For $r\lsim0.05$ it should be replaced with
$|V_{td}V_{ts}|/r$.

The $\delta_{ij}^q$ parameters are further suppressed by the mass
splittings as in Eq.~(\ref{msplit}).
Comparing this to Table \ref{tab:exp}, we find that the
strongest constraint on $r/r_3$ comes
from the bound on $\langle \delta^u_{12} \rangle$. We obtain
 \beq
r/r_3\lsim0.13,
\eeq
in agreement with previous works \cite{Ciuchini:2007cw,Nir:2007ac,Feng:2007ke}.
Estimates for all $\delta_{ij}^q$ parameters
are given in Table \ref{tab:delali} (for $r_3=3$).

\begin{table}[t]
\caption{Upper bounds on the parametric suppression of
   $(\delta_{ij}^{d,u})_{L,R}$ and
   $\langle\delta^{d,u}_{ij}\rangle$ in the hybrid gauge-gravity
   models with alignment and suppressed $\delta^d_{12}$.  For the
   numerical evaluation we take $r/r_3\sim0.13$, $r\leq \hat r \lsim
   1$ and $r_3=3$.  $(\delta_{13,23}^{d})_{L}$ scale as $(3/r_3)$, and
   $\langle\delta^{d}_{13,23}\rangle$ scale as $\sqrt{3/r_3}$.  }
\label{tab:delali}
\begin{center}
\begin{tabular}{ll|ccc} \hline\hline
\rule{0pt}{1.2em}%
$q$ & $ij$\ &  $(\delta^q_{ij})_{L}$ &  $(\delta^q_{ij})_{R}$ &
$\langle\delta^q_{ij}\rangle$ \cr \hline
$d$ & $12$\ & $(r/r_3)|V_{12}|\epsilon_{\rm max}^2\sim0.001$ &
$\frac{(r/r_3) m_d\epsilon_{\rm max}^2}{|V_{12}|m_s}\sim0.001$
& $(r/r_3)\sqrt{m_d/m_s}\epsilon_{\rm max}^2\sim0.001$ \cr
$d$ & $13$\ & $|V_{13}|/r_3\sim0.001$ & $\frac{(r/r_3) m_d\epsilon_{\rm
     max}^2}{|V_{13}|m_b}\sim0.001$
& $\sqrt{r(m_d/m_b)}\ \epsilon_{\rm max}/r_3\sim0.001$ \cr
$d$ & $23$\ & $|V_{23}|/r_3\sim0.01$ & $\frac{(r/r_3) m_s\epsilon_{\rm
     max}^2}{|V_{23}|m_b}\sim0.002$
& $\sqrt{r(m_s/m_b)}\epsilon_{\rm max}/r_3\sim0.006$ \cr
$u$ & $12$\ & $(r/r_3)|V_{12}|\sim0.03$ &
$\frac{(r/r_3) m_u}{|V_{12}|m_c}\sim0.001$
& $(r/r_3)\sqrt{m_u/m_c}\sim0.006$  \cr
\hline\hline
\end{tabular}
\end{center}
\end{table}

We then learn that, in the case that holomorphic zeros play a role in
making the alignment accurate so that the degeneracy is weakest, the
maximal possible effects in the neutral $B_d$, $B_s$ and $D$ systems,
are as follows (for $r_3=3$):

\beq\label{eq:reachhz}
\begin{array}{lc}\label{max}
B_d: & |M_{12}^{\rm susy}/M_{12}^{\rm exp}|\lsim0.0004,\\
B_s: & |M_{12}^{\rm susy}/M_{12}^{\rm exp}|\lsim0.0008,\\
D: & |M_{12}^{\rm susy}/M_{12}^{\rm exp}|\lsim1. \end{array}
\eeq

Thus, the precise alignment further suppresses the new physics effect
in the $B_d$ and $B_s$ mixings. On the other hand, since -- by
construction -- it does not affect the up sector, the milder
degeneracy allows large (and possibly CP violating) effects in the
neutral $D$ system.

%%%%%%%%%%%%%%%%%%%%%%%%%
\section{Probing messengers
\label{sec:messenger}}
We now ask what the constraints derived from FCNC processes,
specifically the upper bound on $r/r_3$ given in
Eq.~(\ref{eq:roverr3}), imply for the parameters of gauge mediation.

\begin{figure}
\includegraphics[scale=1.0]{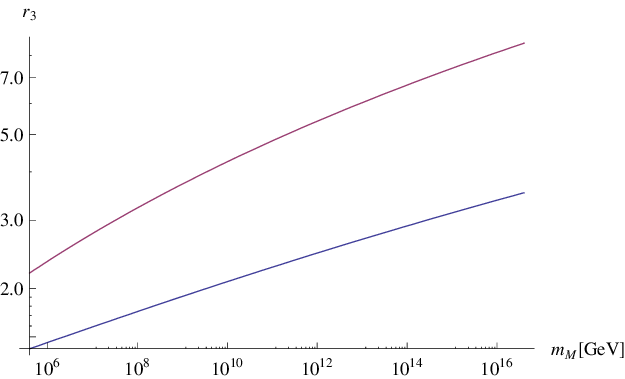}
\caption{\label{fig:r3} The RGE-factor $r_3$ as a function
of the messenger scale for $N_M=1$ (lower curve) and $N_M=3$
(upper curve).}
\end{figure}

Given the soft parameters at the high scale, the RGE-factor $r_3$
defined via Eq.~(\ref{defrthree}) is calculable from the MSSM running
of the soft squark masses, the one loop running of which is also
discussed in Appendix \ref{app:rge}.  Neglecting contributions from
the electroweak gauge couplings, one obtains an analytical expression
for $r_3$ (see, {\it e.g.}, \cite{Martin:1997ns}):
\beq \label{eq:r3-exact}
r_3 =r_3(m_M)=1 + \frac{8}{3 \pi} \left( \int_{\ln( m_Z)}^{\ln(m_M)}
dt  \frac{\alpha_3^3(t)}{\alpha_3^2(m_M)} \right)
\frac{M_3^2(m_M)}{ \tilde m_{12}^2(m_M)}  .
\eeq
Here, $M_3$ denotes the gluino mass and $\tilde m_{12}^2$ is defined
in Eq.~(\ref{eq:m12tilde}). In messenger models of gauge mediation,
the ratio $M_3^2/\tilde m_{12}^2$ is determined by a simple formula at
the scale of mediation:
\beq \label{eq:M3mtilde}
\frac{M_3^2(m_M)}{ \tilde m^2_{12}(m_M)}=\frac{3}{8}N_M  +
{\cal{O}}\left[\left(\frac{\alpha_i}{\alpha_3}\right)^2\right]  , ~~
i=1,2 , ~~\mathrm{for} ~~ q=Q_L,U_R,D_R ,
\eeq
where $N_M$ denotes the number of color-triplet messengers.  We
explicitly see that in our approximation, due to the universality of
the initial conditions and the running, $r_3$ is universal for $Q_L,
U_R$ and $D_R$ soft masses.  We depict $r_3$ as a function of the
messenger scale for $N_M=1$ and $N_M=3$ in Fig.~\ref{fig:r3}. It
depends logarithmically on $m_M$, and grows with $N_M$.

The parameter $r$ introduced in the initial conditions of
gauge-gravity models at the messenger scale $m_M$,
Eq.~(\ref{inicon}), can be expressed as a ratio of soft squark masses:
\beq\label{eq:r-def}
r = \frac{ \tilde m_{12-{\rm gravity}}^2}{\tilde m_{12-{\rm gauge}}^2}
\sim \left( \frac{m_M}{m_{\rm Pl}} \right)^2 \left( \frac{4
    \pi}{\alpha_3(m_M)} \right )^2 \frac{3}{8} \frac{1}{N_M} ,
\eeq
where $m_{\rm Pl} \sim 10^{19} \, \mbox{GeV}$ denotes the Planck mass.
In Eq.~(\ref{eq:r-def}) we again neglect
contributions other than from the strong interaction as well as running
of the gravity-induced soft terms above $m_M$.

\begin{figure}
\includegraphics[scale=1.0]{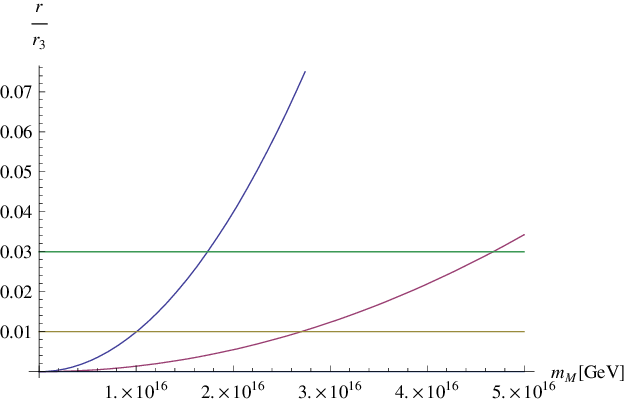}
\caption{\label{fig:roverr3} $r/r_3$ as a function of the messenger
  scale for $N_M=1$ (upper curve) and $N_M=3$ (lower curve)
from Eq.~(\ref{eq:r-def}). The
  horizontal lines correspond to the FCNC upper bounds of
  Eq.~(\ref{eq:roverr3}).}
\end{figure}

Eq.~(\ref{eq:roverr3}) implies the existence of an upper bound on the
messenger scale or, in other words, a minimal separation between the
scales of gravity- and gauge-mediation. We find that flavor physics
determines this to be about three orders of magnitude, {\it i.e.},
$m_M \lesssim m_{\rm Pl}/10^3$. A larger number of messengers gives a
heavier spectrum, and hence a weaker bound.
This is also illustrated in Fig.~\ref{fig:roverr3}.

In writing Eq.~(\ref{eq:r-def})
we assumed that the highest $F$-term contributes to gauge mediation.
If this is not the case, $r$ gets enhanced by
$\langle F \rangle^2/\langle F_M\rangle^2$, the square of the ratio
of the highest $F$-term vev to the one that couples to the messengers.
The flavor constraint Eq.~(\ref{eq:roverr3}) requires then a low $m_M$, or,
turning the argument around, indicates gravity-mediated contributions can be
non-negligible even if the scale of gauge mediation is low.

It has been pointed out recently that hidden sector effects modify in
general the initial conditions below which the known MSSM-RG equations
apply \cite{Cohen:2006qc}. If the hidden sector is weakly interacting,
then the effects are small and our analysis holds to this degree.  If
the renormalization is non-perturbative, our analysis will depend on
the unknown hidden sector physics. A general framework, termed general
gauge mediation, to account for this has been outlined in
\cite{Meade:2008wd}.

Within general gauge mediation, our analysis is affected in the
following ways:
\begin{enumerate}
\item The relation between the gluino mass and the soft squark masses,
Eq.~(\ref{eq:M3mtilde}), can receive order one corrections. The
outcome of this for the example of a change of factor three in the
initial conditions is illustrated by the difference in the curves of
Fig.~\ref{fig:r3}. In other words, we cannot calculate
$r_3$ without knowledge of the hidden sector.
\item The initial conditions for the soft squark masses,
Eq.~(\ref{inicon}), are not of perturbative messenger gauge-mediation
type. In particular, the soft masses for $Q_L,U_R$ and $D_R$ are
renormalized differently, and in general we need to introduce several
RGE-parameters $r_3$. Note that, as in the minimal case, in the limit of
$\alpha_1, \alpha_2 \to 0$ we recover  universality of soft masses and
hence, of $r_3$. Since the corrections arise in full generality
non-perturbatively, this might not be representing the true spectrum.
\item Unlike perturbative messenger mediation, general gauge mediation
does not exclude $\tilde m_{12}^2(m_M) <0$. Consequently, $r_3 <1$
becomes possible, see Eq.~(\ref{eq:r3-exact}). To avoid a tachyonic
spectrum, then, however, a very large RGE effect is required such that
$r_3 <0$.
\item We cannot express $r$ in terms of messenger parameters
as simply as Eq.~(\ref{eq:r-def}).
\end{enumerate}

What, however, still remains valid in general gauge mediation is the
form of Eq.~(\ref{inicon}). In particular, the hidden sector effects
do not introduce further flavor violation into the soft masses because
gauge mediation respects the $U(3)^5$ global flavor symmetry.

By not fixing $r_3$ to a specific, minimal gauge-mediation value, we
have hence mimicked hidden sector effects in Section \ref{sec:phe}.

%%%%%%%%%%%%%%%%%%%%%%%%%
\section{Conclusions}
\label{sec:con}
We considered supersymmetric models where squark masses are dominated
by gauge-mediated contributions, yet gravity-mediated contributions
are not negligible. Such a situation arises when the messenger scale
is not much below $\alpha_3 m_{\rm Pl}$, or when the $F$-term that
leads to gauge mediation is at a scale much lower than the highest
$F$-term. We further assumed that the gravity-mediated contributions
follow selection rules that arise from a Froggatt-Nielsen symmetry that
explains the hierarchy in the Yukawa couplings. Such models constitute
an example of viable and natural supersymmetric models that are not
minimally flavor violating (non-MFV). The mass splittings and flavor
decomposition of sfermions can perhaps be directly measured in the
ATLAS/CMS experiments \cite{Feng:2007ke}.

We posed here the question of whether measurements of FCNC processes,
such as neutral meson mixing, can show
signals of such non-MFV models. We found that the strongest bound on
the mass splitting between the first two squark generations
$\Delta \tilde m_{12}^2/\tilde m_{12}^2$ comes from
$K^0-\overline{K}^0$ mixing, and is of ${\cal O}(0.03)$. This splitting
reflects the relative size of the gravity- and gauge-mediated
contributions which, at the mediation scale, gets lifted by an inverse
RGE-factor w.r.t.~the physical splitting at the electroweak scale.
We obtain for the respective splitting at the mediation scale
a value that is constrained to be
below  ${\cal O}(0.1)$ for minimal gauge mediation with one messenger,
or even as large as  ${\cal O}(0.3)$ in general gauge mediation, or with
several messengers.

The slepton sector has also been studied within hybrid
gauge-gravity mediation \cite{Feng:2007ke}.
Assuming the simplest FN charge assignments, no parametric suppression of the
1-2 lepton mixing angle and minimal gauge mediation giving sleptons
lighter than squarks, the bounds from lepton flavor changing processes
on the splittings are stronger than those from the quarks.

Given the constraint on the splitting, the order of magnitude
predictions that follow from the FN symmetry, and the RGE effects, we
evaluated the maximal possible modifications to the Standard Model
predictions to various FCNC processes. We found that the effects on
the $B_d-\overline{B}_d$ and $B_s-\overline{B}_s$ mixing amplitudes is
generically below the percent level, but can be of order ten percent for large
$\tan \beta$.
 It is maximized when the RGE suppression is minimal.

On the other hand, the effect on the $D^0-\overline{D}^0$ mixing amplitude
can be ${\cal O}(1)$ (and CP violating), though in the simplest models it
is at most of order five percent. We found also that the ratio of
$B_s \to \mu^+ \mu^-$ to $B_d \to \mu^+ \mu^-$ branching ratios
is sensitive to the FN flavor symmetries.

Further possibilities to test FN gravity, that is, Planck scale physics,
with rare decays are pointed out. Particularly promising are searches
for right-handed currents, if possible even in conjunction with
CP-violation.

When thinking about the future of experimental flavor physics, and
evaluating the sensitivity to new physics of, for example, a super-B
factory \cite{superb,Buchalla:2008jp}, a question that often arises is
the following: What experimental accuracy is worth achieving, given
well-motivated models
of new physics as well as theoretical (QCD-related) uncertainties.
Eqs.~(\ref{eq:FN-reach}), (\ref{eq:FN-reach-largetb})
and (\ref{eq:reachhz}) provide a concrete answer
-- within a specific but well-motivated and natural framework -- to
this question. An accuracy of order a few percent in measurements
related to neutral $D$, $B_d$ or $B_s$ mixing may be sensitive to new
physics. Since the new physics that we discuss introduces, in general,
new CP violating phases of order one, a theoretically clean signal for
the new physics can be established by measuring CP asymmetries at that
level.

%%%%%%%%%%%%%%%
\acknowledgments
This work was supported by a grant from the G.I.F., the
German--Israeli Foundation for Scientific Research and Development,
and by the Minerva Foundation. The work of YN is supported in part
by the United States-Israel Binational Science Foundation (BSF),
and by the Israel Science Foundation (ISF).

%%%%%%%%%%%%%
\appendix
\section{RGE effects}
\label{app:rge}
In the Appendix, we present the renormalization group equations for
the quark and squark parameters relevant to our framework.
(General formulae are given in Ref. \cite{Martin:1993zk}.)
We use the following approximations:
\begin{enumerate}
   \item We neglect the RGE effects of the first and second generation
     Yukawa couplings $y_u,y_d,y_s$ and $y_c$.
     \item We neglect the RGE effects that involve
       $|V_{ts}|^2,|V_{td}|^2$ and $V_{td}V_{ts}^*$.
       \item We neglect the effects of the off-diagonal elements in the
         squark mass-squared matrices on the running of the diagonal
         terms.
     \end{enumerate}
(Within the special class of models discussed in Section \ref{sec:hol},
some of these approximations are not valid, and then we do include
the relevant factors.)

We obtain for the CKM mixing angles \cite{Naculich:1993ah}
\beq\label{ckmrge}
16\pi^2\frac{d}{dt}\ln V_{\alpha\beta}=\left\{
   \begin{array}{ccl}
     -y_t^2-y_b^2 & \mbox{for}& ~ V_{ub},V_{cb},V_{td},V_{ts}\\
     0 & \mbox{for}&~ V_{ud},V_{us},V_{cd},V_{cs},V_{tb} \end{array} \right.
\eeq
and for the Yukawa coupling ratios (or, equivalently, mass ratios)
\beqa\label{rgflck}
16\pi^2\frac{d}{dt}\ln (y_u/y_c)&=&0,\no\\
16\pi^2\frac{d}{dt}\ln (y_c/y_t)&=&-3y_t^2-y_b^2,\no\\
16\pi^2\frac{d}{dt}\ln (y_d/y_s)&=&0,\no\\
16\pi^2\frac{d}{dt}\ln (y_s/y_b)&=&-y_t^2-3y_b^2,\no\\
16\pi^2\frac{d}{dt}\ln [V_{cb}/(y_c/y_t)]&=&2y_t^2,\no\\
16\pi^2\frac{d}{dt}\ln [V_{cb}/(y_s/y_b)]&=&2y_b^2.
\eeqa

For the diagonal elements in the soft squark mass-squared matrices, we
obtain ($i=1,2,3$)
\beqa\label{diarge}
16\pi^2\frac{d}{dt}(M^2_{\tilde Q_L})_{ii}&=&2[(M^2_{\tilde Q_L})_{33}
+(M_{\tilde U_R}^2)_{33}+m_{H_u}^2]y_t^2\delta_{i3}\no\\
&+&2[(M^2_{\tilde Q_L})_{33}+(M_{\tilde D_R}^2)_{33}+m_{H_d}^2]y_b^2\delta_{i3}
-\frac{32}{3}g_3^2|M_3|^2+{\cal O}(g_2^2,g_1^2),\no\\
16\pi^2\frac{d}{dt}(M^2_{\tilde U_R})_{ii}&=&4[(M^2_{\tilde U_R})_{33}
+(M_{\tilde Q_L}^2)_{33}+m_{H_u}^2]y_t^2\delta_{i3}
-\frac{32}{3}g_3^2|M_3|^2+{\cal O}(g_2^2,g_1^2),\no\\
16\pi^2\frac{d}{dt}(M^2_{\tilde D_R})_{ii}&=&4[(M^2_{\tilde D_R})_{33}
+(M^2_{\tilde Q_L})_{33}+m_{H_d}^2]y_b^2\delta_{i3}
-\frac{32}{3}g_3^2|M_3|^2+{\cal O}(g_2^2,g_1^2).
\eeqa
For the off-diagonal terms involving the third generation, we obtain,
in the super-CKM basis (where gluino couplings and quark masses are
diagonal), ($i \neq 3$)
\beqa\label{offdiarge}
16\pi^2\frac{d}{dt}(\widetilde M^2_{\tilde U_L})_{i3}&=&[(M^2_{\tilde
   Q_L})_{ii}+(M^2_{\tilde Q_L})_{33} +2(M_{\tilde D_R}^2)_{33}+2m_{H_d}^2]
y_b^2V_{ib}V_{tb}^*
+(y_t^2+y_b^2)(\widetilde M^2_{\tilde U_L})_{i3}\no\\
16\pi^2\frac{d}{dt}(\widetilde M^2_{\tilde D_L})_{i3}&=&[(M^2_{\tilde
   Q_L})_{ii}+(M^2_{\tilde Q_L})_{33} +2(M_{\tilde U_R}^2)_{33}+2m_{H_u}^2]
y_t^2V_{ti}^*V_{tb}+(y_t^2+y_b^2)(\widetilde M^2_{\tilde D_L})_{i3}\no\\
16\pi^2\frac{d}{dt}(\widetilde M^2_{\tilde U_R})_{i3}&=&
2y_t^2(\widetilde M^2_{\tilde U_R})_{i3},\no\\
16\pi^2\frac{d}{dt}(\widetilde M^2_{\tilde D_R})_{i3}&=&
2y_b^2(\widetilde M^2_{\tilde D_R})_{i3}.
\eeqa
The $1-2$ terms are, within our approximations, RGE invariant:
\beq
16\pi^2\frac{d}{dt}(M^2_{\tilde Q_L,\tilde U_R,\tilde D_R})_{12}=0.
\eeq

%%%%%%%%%%%%%
\subsection{$(\delta^{q}_{12})_A$}
With our approximations (which hold much more generally
than within our specific framework), almost all parameters related to
just the first two generations, and, in particular,
\beq\label{onetwoinv}
(M^2_{\tilde q_A})_{12},\ \ \
(M^2_{\tilde q_A})_{22}-(M^2_{\tilde q_A})_{11},
\eeq
are RGE invariant.
In models (as ours) where $|(M^2_{\tilde q_A})_{12}|\ll
|(M^2_{\tilde q_A})_{22}-(M^2_{\tilde q_A})_{11}|$,
Eq.~(\ref{onetwoinv}) further implies the RGE invariance of
\beq\label{tonetwoinv}
(\tilde V^q_A)_{12},\ \ \ \Delta\tilde m^2_{q_{A2}q_{A1}}.
\eeq
The only parameter related to the first two generations which
is not RGE invariant is the average squark mass. The universal
QCD effect on the running of the diagonal mass-squared terms
is actually the only RGE effect that (for running from high scale,
as in our framework) can be significantly larger than one.
This is taken into account by the factor $r_3$ defined in
Eq.~(\ref{defrthree}). Numerical values within gauge mediation
are discussed in Section \ref{sec:messenger}.

The parameters of interest for our purposes are the $(\delta^{q}_{ij})_A$
parameters. We analyze the RGE implications on these parameters using
the two generation approximation of Eq.~(\ref{twogendel}). {}From
Eqs.~(\ref{rgflck},\ref{tonetwoinv},\ref{defrthree}) we learn that
\beq
(\delta^{q}_{12})_A(\mu=m_Z)=\frac{1}{r_3}(\delta^{q}_{12})_A(\mu=m_M).
\eeq
Within our framework, where the structures of the quark and squark mass
matrices are related by the FN symmetry, this leads to the values of
the $(\delta^q_{12})_L$ as given in Eq. (\ref{eq:delql}) and
$(\delta^q_{12})_R$  as given in Eq. (\ref{eq:delqr}).

%%%%%%%%%%%%%
\subsection{$(\delta^{q}_{i3})_R$}
Within our approximation, we also find from Eqs.~(\ref{rgflck}) and
(\ref{offdiarge}) that the following two combinations
of squark and quark parameters are RGE invariant:
\beq\label{delrit}
\frac{(\widetilde M^2_{\tilde U_R})_{i3}}{(y_{u_i}/y_t)/|V_{cb}|},\ \ \
\frac{(\widetilde M^2_{\tilde D_R})_{i3}}{(y_{d_i}/y_b)/|V_{cb}|}\ \ \ (i=1,2).
\eeq
The RGE effects on the splittings are as follows (see
Eq.~(\ref{diarge})):
\beqa
 \label{eq:qRsplitting}
[(M^2_{\tilde U_R})_{33}-(M^2_{\tilde U_R})_{ii}](\mu=m_Z)
&\sim&\tilde m^2_q,\no\\
\ [(M^2_{\tilde D_R})_{33}-(M^2_{\tilde D_R})_{ii}](\mu=m_Z)
&\sim&\hat r\tilde m^2_q.
\eeqa
These equations lead to the estimates of $(\tilde V^q_R)_{i3}$ given
in Eq. (\ref{eq:susyk}), $(K^q_R)_{i3}$ as given in Eq. (\ref{K1}),
and $(\delta^q_{i3})_R$ as given in Eq. (\ref{eq:delqr}).

%%%%%%%%%%%%
\subsection{$(\delta^{q}_{i3})_L$}
The situation regarding $(\delta^{q}_{i3})_L$ is less simple than the other
cases. Here, Eqs.~(\ref{ckmrge}) and (\ref{offdiarge}) imply,
unlike the analogous case for $\tilde q_R$ (see Eq.~(\ref{delrit})),
that $(M^2_{\tilde Q_L})_{i3}/|V_{ib}|$ is not RGE invariant. Consider first
the $\tilde U_L$ sector, and assume for simplicity small $\tan\beta$
(so that the  $y_b^2$-dependent terms in Eq. (\ref{offdiarge}) can be neglected):
\beq
16\pi^2\frac{d}{dt}\ln\frac{(M^2_{\tilde
     U_L})_{i3}}{|V_{ib}|}=2(y_t^2+y_b^2).
\eeq
For the relevant mass-squared difference, we obtain from
Eq.~(\ref{diarge}):
\beqa\label{diargeb}
16\pi^2\frac{d}{dt}[(M^2_{\tilde Q_L})_{33}-(M^2_{\tilde
  Q_L})_{ii}]&=&2[(M^2_{\tilde Q_L})_{33}
+(M_{\tilde U_R}^2)_{33}+m_{H_u}^2]y_t^2\no\\
&+&2[(M^2_{\tilde Q_L})_{33}+(M_{\tilde D_R}^2)_{33}+m_{H_d}^2]y_b^2.
\eeqa
The conclusion is that the RGE effects on both $(\tilde V^u_L)_{i3}$
and on $|V_{ib}|$ are ${\cal O}(1)$ and different from each other. Yet,
at low energy, in the up quark mass basis, we have (recall
Eq.~(\ref{eq:susyk}) is in the FN basis)
\beq
|(\tilde V^u_L)_{i3}| \sim r|V_{ib}|.
\eeq
When the MFV $y_b^2$ dependent terms are taken into account, we obtain
Eq. (\ref{eq:susyk}) for $(\tilde V_L^u)_{i3}$,
Eq. (\ref{K1}) for $(K^u_L)_{i3}$, and
Eq. (\ref{eq:delql}) for $(\delta^u_{i3})_L$.

Next consider the running of $(M^2_{\tilde D_L})_{i3}$ in the down
quark mass basis. The second term on the right hand side of the
relevant Eq.~(\ref{offdiarge}) is smaller by a factor of ${\cal O}(r)$
than the first and so $|(\tilde V^d_L)_{i3}|\approx |V_{ti}|$.
Eqs. (\ref{K1}) for $(K^d_L)_{i3}$, and (\ref{eq:delql}) for $(\delta^d_{i3})_L$
follow.

%%%%%%%%%%%%%%%%%%%%%

\end{document}